\documentclass[aps,pra,twocolumn,superscriptaddress]{revtex4}

\usepackage[T1]{fontenc} 

\usepackage{amsmath,amssymb}
\usepackage{psfrag}
\usepackage[dvips]{epsfig}
\usepackage{epsfig}

\usepackage{amsfonts}
\usepackage{graphicx}
\usepackage{rotating}

\DeclareMathOperator{\trace}{Tr}
\newcommand{\tracep}[1]{\trace\left[ #1 \right]}

\newcommand{\half}{\frac{1}{2}}

\newcommand{\parenth}[1]{\left( #1 \right)}

\newcommand{\bal}{\begin{align}}
\newcommand{\eal}{\end{align}}

\newcommand{\rhof}{\rho_{\rm f}}
\newcommand{\Uf}{U} 

\newcommand{\comm}[2]{\ensuremath{\left[#1, #2\right]}}
\newcommand{\evol}[2][]{\mathcal{E}_{#1} \left\{ #2 \right\}}

\newcommand{\be}{\begin{equation}}
\newcommand{\ee}{\end{equation}}

\newcommand{\ev}[1]{\langle #1 \rangle}

\newcommand{\ie}{{\rm i.e.}}
\newcommand{\eg}{{\rm e.g.}}

\newcommand{\TKK}{Laboratory of Physics, Helsinki University of Technology
P. O. Box 4100, 02015 TKK, Finland}
\newcommand{\LTL}{Low Temperature Laboratory, Helsinki University of
Technology, P.O. Box 3500, 02015 TKK, Finland}
\newcommand{\BERKELEY}{Department of Chemistry and Pitzer Center for
Theoretical Chemistry, University of California, Berkeley, CA 94720}

\begin{document}
\title{Suppression of $1/f^\alpha$ noise in one-qubit systems}

\author{Pekko~Kuopanportti}\email{pekko.kuopanportti@tkk.fi}\affiliation{\TKK}
\author{Mikko~M\"ott\"onen}\affiliation{\TKK}\affiliation{\LTL}
\author{Ville~Bergholm}\affiliation{\TKK}
\author{Olli-Pentti~Saira}\affiliation{\TKK}\affiliation{\LTL}
\author{Jun~Zhang}\affiliation{\BERKELEY}
\author{K.~Birgitta~Whaley}\affiliation{\BERKELEY}

\date{\today}

\begin{abstract}
  We investigate the generation of quantum operations for one-qubit
  systems under classical noise with $1/f^\alpha$ power spectrum, where
  $2>\alpha > 0$.  We present an efficient way to approximate the
  noise with a discrete multi-state Markovian fluctuator.
  With this method, the average temporal evolution of the qubit density matrix
  under $1/f^\alpha$ noise can be feasibly determined from recently
  derived deterministic master equations. We obtain qubit
  operations such as quantum memory and the {\sc NOT} gate to high fidelity by a gradient based optimization algorithm.
   For the {\sc NOT} gate, the
  computed fidelities are qualitatively similar to those obtained
  earlier for random telegraph noise. In the case of quantum memory
  however, we observe a nonmonotonic dependency of the fidelity on the
  operation time, yielding a natural access rate of the memory.
\end{abstract}

\maketitle

\section{Introduction}
\label{sc:intro} In solid-state realization of qubits, material specific fluctuations typically induce the major contribution to the intrinsic
noise. Much effort has been focused on the preservation of the state in a quantum memory in the presence of $1/f^\alpha$ noise since this is a
ubiquitous form of noise encountered in solid-state qubit applications~\cite{faoroviola,Paladino2002,Falci2004}. Both charge and spin qubits are
susceptible to noise of this form. For Josephson junctions, both charge noise~\cite{astafiev06,astafiev04} and critical current
noise~\cite{wellstood04,muck05}
 have been measured to have $1/f^\alpha$ power spectral densities.
 Similar charge fluctuations are responsible for the well-known $1/f^\alpha$ nature of low frequency noise in
single electron transistors~\cite{zimmerli92}.
 Background charge fluctuations
resulting in $1/f^\alpha$ noise spectra are considered to be the most important source of dephasing in Josephson junction
qubits~\cite{astafiev06,astafiev04,nakamura}. Spin qubits such as those formed from donor spins in semiconductors are susceptible to nuclear
spin noise deriving from dipolar coupling between environmental nuclear spins. The nuclear spin bath couples to the donor spins by hyperfine
interactions, which renders the dynamics of the nuclear spins to cause dephasing. Recent calculations for a phosphorus donor in silicon show
that the high frequency component of the nuclear spin noise is approximately described by a $1/f^\alpha$ power spectrum~\cite{deSousa06}.
Electron spin qubits implanted into silicon~\cite{Schenkel06} are also affected by relaxation of dangling bonds deriving from oxygen vacancies
at the Si/SiO$_2$ interface. This gives rise to a magnetic noise with a $1/f^\alpha$ spectrum that is the dominant mechanism for phase
fluctuations of donor spins near the surface~\cite{deSousa07}. Another form of noise closely related to $1/f^\alpha$ noise is random telegraph
noise (RTN), which arises from coupling of individual bistable fluctuators to a qubit~\cite{deSousa05,Nakamura2002, Galperin03, Paladino2002,
Savo1987, Wakai1987, Fujisawa2000, Kurdak1997}.

Several approaches to suppress decoherence based on pulse design have been proposed in the literature. Among them, dynamical decoupling schemes
average out the unwanted effects of the environmental interaction through the application of suitable control pulses~\cite{violalloyd,
violalloydknill}. Application of these schemes often involves hard pulses with instantaneous switchings and unbounded control amplitudes,
resulting in a range of validity restricted to time scales for which the pulse duration is much less than the noise correlation
time~\cite{Kofman01, Kofman04}.  In Ref.~\cite{hifid}, a direct pulse optimization method restricted to bounded control pulses was developed for
implementing one-qubit operations in a noisy environment. This initial work on noise suppression addressed the example of a single qubit system
under the influence of classically modeled random telegraph noise, such as might arise from a single bistable fluctuator.

In this paper, we extend the work of Ref.~\cite{hifid} to the physically relevant situation of $1/f^\alpha$ noise where $2>\alpha > 0$. This
kind of noise is known to result, for example, from a set of bistable fluctuators~\cite{Weissman88, Paladino02, kaulakys, Galperin06}, \ie, RTN
sources. We investigate two ways to approximate the $1/f^\alpha$ noise for computer simulations, namely, the sum of independent RTN fluctuators
and a single discrete multi-state Markovian noise source. We show that the single fluctuator provides a much more efficient way to model
$1/f^\alpha$ noise than independent RTN fluctuators. Furthermore, the average temporal evolution of the density matrix under this Markovian
noise can be exactly described by a set of deterministic master equations derived in Ref.~\cite{equdyn}. Using this approach, we avoid the heavy
computational task arising from the numerical evaluation of the density matrix averaged over a large number of different sample paths of the
noise as computed in Ref.~\cite{hifid}. This framework will not only significantly accelerate the convergence of the control pulse sequence
optimization, but also allows further theoretical analysis. Using these master equations, we employ gradient based optimization procedures to
obtain pulse sequences that suppress $1/f^\alpha$ noise for quantum memory and for a {\sc NOT} gate. Comparisons with composite pulses designed
to eliminate systematic errors and with refocusing pulses demonstrate that the numerically optimized pulse sequences yield the highest
fidelities.

The remainder of this paper is organized as follows. In Sec.~\ref{sc:model}, we show how to efficiently approximate the $1/f^\alpha$ noise by a
multi-state Markovian fluctuator. In Sec.~\ref{sc:dyna}, we define the fidelity of qubit operations, review the master equations describing the
average evolution of the qubit density matrix in the presence of the noise and describe the numerial optimization procedure.
Sections~\ref{sc:results} and~\ref{sc:resultsNOT} present optimized control pulse sequences and the achieved fidelities for quantum memory and
for the {\sc NOT} gate, respectively. Finally, Sec.~\ref{sc:discussion} concludes and indicates further applications of the method.

\section{One-qubit system subject to $1/f^\alpha$ noise}
\label{sc:model}

We consider a one-qubit system described by the effective Hamiltonian
\begin{equation}
\label{eq:qubham}
H=\half a(t)\sigma_x+\half\eta(t)\sigma_z,
\end{equation}
where $a(t)\in [-a_{\rm max},a_{\rm max}]$ is the external control field applied along the $x$ direction and $\eta(t)$ is the classical noise
signal perturbing the system along the $z$ direction. The noise source $\eta(t)$ can be characterized by its autocorrelation function
\begin{equation}
\label{eq:autocorr}
C(t) \equiv \ev{\eta(0)\eta(t)}
= \lim_{T\to\infty}\frac{1}{T}\int_{-T/2}^{T/2} \eta(s)\eta(s+t) \: \textrm{d}s,
\end{equation}
the Fourier transformation of which defines the noise power
spectral density as
\begin{equation}
\label{eq:psdvscorr} S(f) = \int_{-\infty}^\infty C(t) e^{-i 2 \pi f t} \textrm{d}t.
\end{equation}
For a single RTN source with the amplitude $\Delta$ and correlation time $\tau_{\rm c}$, the autocorrelation function is given by~\cite{kirton}
\begin{equation}
\label{eq:rtnautocorr}
C_\text{RTN}(t) = \Delta^2 e^{-2 |t|/\tau_{\rm c}},
\end{equation}
and the corresponding power spectral density by
\begin{equation}
\label{eq:rtnpsd}
S_\text{RTN}(f)=\frac{\Delta^2 \tau_{\rm c}}{1+(\pi f \tau_{\rm c})^2}.
\end{equation}

A standard way to simulate $1/f^\alpha$ noise is to use an ensemble of
$K$ independent uncorrelated RTN processes~\cite{faoroviola,
  Weissman88, kaulakys}.  Let $\eta_k(t)$ be a symmetric RTN signal
switching between values $-\Delta_k$ and $\Delta_k$ with the
correlation time $\tau_k \equiv 1/\gamma_k$, where $\gamma_k$ is the
transition rate between the two states. The total noise process
appears in the Hamiltonian (\ref{eq:qubham}) as $\eta(t) =\sum_{k=1}^K
\eta_k(t)$. Since the RTN sources are independent,
Eqs.~(\ref{eq:autocorr}) and (\ref{eq:rtnautocorr}) yield the
autocorrelation function
\begin{equation}
\label{eq:rtnsumautocorr}
C(t) = \sum_{k=1}^K\Delta_k^2
e^{-2|t|/\tau_k}= \sum_{k=1}^K\Delta_k^2 e^{-2\gamma_k |t|},
\end{equation} and
the corresponding power spectral density is given by
\begin{equation}
\label{eq:rtnsumpsd}
S(f) = \sum_{k=1}^K \frac{\Delta_k^2 \gamma_k}{\gamma_k^2+(\pi f)^2}.
\end{equation}
Introducing the density of transition rates $g(\gamma)$ and expressing the noise strength $\Delta$ as a function of the transition rate, we can
replace the summation in Eq.~(\ref{eq:rtnsumpsd}) by an integration, which yields
\begin{equation}
\label{eq:rtnpsdintegral}
S(f)= \int_{\gamma_{\rm min}}^{\gamma_{\rm max}} \frac{
  \Delta^2(\gamma)g(\gamma)\gamma}{\gamma^2+(\pi f)^2} \: \textrm{d}\gamma,
\end{equation}
where $\gamma_{\rm min}$ and $\gamma_{\rm max}$ are minimal and
maximal transition rates, respectively. Provided that
\begin{equation}
  \label{eq:1}
  \Delta^{2}(\gamma)g(\gamma) = 2 A/\gamma,
\end{equation}
where $A$ is a constant, the power spectral density in
Eq.~(\ref{eq:rtnpsdintegral}) becomes~\cite{kaulakys}
\begin{align}
\label{eq:rtnsumpsd2}
S(f) &= \frac{2 A}{\pi f}\left[\arctan \parenth{\frac{\gamma_{\rm max}}{\pi f}}
-\arctan \parenth{\frac{\gamma_{\rm min}}{\pi f}}\right] \notag\\
&\simeq \frac{A}{f},\quad \gamma_{\rm min} \ll \pi f \ll \gamma_{\rm max}.
\end{align}
Thus Eq.~\eqref{eq:rtnsumpsd2} yields an approximation to the $1/f$ power spectrum. To generate a general $1/f^\alpha$ power spectral density
for $2>\alpha >  0$, we can choose
\begin{equation}
  \label{eq:2}
  \Delta^{2}(\gamma)g(\gamma) = 2 A\gamma^{-\alpha}
\end{equation}
as shown in~\cite{kaulakys}.

Although the above method yields a valid approximation for the
$1/f^\alpha$ spectrum, it is computationally inefficient. In
particular, the number of distinct noise states increases
exponentially with the number of RTN fluctuators~$K$, \ie, the number
of terms in the sum of Eq.~\eqref{eq:rtnsumpsd} approximating the
$1/f^\alpha$ noise. Since the size of the differential equation system
describing the average qubit dynamics increases linearly with the
number of noise states~\cite{equdyn}, in practice one has to restrict
the computation to a rather small number of independent RTN
fluctuators.

To overcome this problem, we present a conceptually different way of generating the desired $1/f^\alpha$ noise spectrum using a single
multi-state Markovian fluctuator. Consider a continuous-time Markovian noise process with $M$~discrete noise states. Let $\Gamma_{kj}$ denote
the transition rate from the $j$th state to the $k$th one. In order to preserve total probability, we must have
\begin{equation}
\label{eq:probconserv}
\sum_{j=1}^M \Gamma_{jk}=0 \quad \text{for all} \; k=1,2,\dots,M.
\end{equation}

Let us assume that the transition rates are symmetric, \ie, $\Gamma = \Gamma^T$. Under this assumption the noise process has a steady-state
solution in which the different noise states are equally probable. In order for the noise to be unbiased, \ie, $\ev{\eta}=0$, the amplitudes
$b_k$ associated with the noise states must satisfy \be \label{eq:amplitudes} \sum_{k=1}^M b_k=0. \ee Thus the autocorrelation is given by
\begin{equation}
\label{eq:markovcorr}
C(t) = \ev{\eta(t)\eta(0)} = \frac{1}{M} b^T e^{\Gamma |t|} b.
\end{equation}
Since $\Gamma$ is symmetric, we can diagonalize it with an orthogonal matrix $V$ as $\Gamma = V \Lambda V^T$, where the real diagonal matrix
$\Lambda =\textrm{diag}\{\lambda_k\}_{k=1}^M$ carries the eigenvalues of~$\Gamma$ in a descending order. Defining $\chi := \frac{1}{\sqrt{M}}
V^T b$, we rewrite Eq.~(\ref{eq:markovcorr}) in the form of Eq.~\eqref{eq:rtnsumautocorr} as
\begin{equation}
\label{eq:markovcorr2}
C(t) = \chi^T e^{\Lambda |t|} \chi =\sum_{k=1}^M \chi_k^2 e^{\lambda_k |t|}.
\end{equation}
In order to use this multi-state Markovian fluctuator to approximate $1/f^\alpha$ noise, we have to choose the eigenvalues $\lambda_k$ and the
amplitudes $\chi_k$ such that Eq.~\eqref{eq:2} is fulfilled. Moreover, we must construct the orthogonal matrix~$V$ such that $\Gamma=V\Lambda
V^T$ satisfies Eq.~\eqref{eq:probconserv}, the amplitudes~$b_k$ satisfy Eq.~\eqref{eq:amplitudes}, and the off-diagonal elements of~$\Gamma$
must be non-negative.

One way to satisfy these requirements is to pick an integer $m \ge 2$ and set $M=2^m$ and to choose the eigenvalues as
\begin{equation*}
\{\lambda_k\}_{k=1}^M = -2 \{0, \gamma_{\min}, \gamma_{\min}+\delta,
\gamma_{\min}+2\delta, \ldots, \gamma_{\max} \},
\end{equation*}
where $\gamma_{\max}=(M-2)\delta+\gamma_{\min}$ and $0<\delta\leq\gamma_{\min}$. Hence, the distribution of the transition rates $g(\gamma)$ is
uniform on $[\gamma_{\min}, \gamma_{\max}]$. Then we set $V=H^{\otimes m}$, where $H$ is the Hadamard matrix
\begin{eqnarray*}
H=\frac{1}{\sqrt{2}}\left(\begin{matrix}1&1\\1& -1
  \end{matrix}\right).
\end{eqnarray*}
Explicit calculation shows that these choices ensure that Eq.~\eqref{eq:probconserv} is satisfied.  To fulfill Eqs.~\eqref{eq:2}
and~\eqref{eq:amplitudes}, we set $\chi_1=0$ and $\chi_k=\gamma_k^{-\alpha/2}$ for $k=2$, \dots, $M$, where $\gamma_k$ is equal to
$\gamma_{\min}+(k-2)\delta$.  It can be shown that this construction will also produce transition matrices~$\Gamma$ with non-negative
off-diagonal elements.  Hence we have provided an efficient way to implement $1/f^\alpha$ noise.  Note that the $M$-state Markovian fluctuator,
Eq.~\eqref{eq:markovcorr2}, corresponds formally to Eq.~\eqref{eq:rtnsumautocorr} with $M-1$ non-vanishing RTN fluctuators.  Thus we have
achieved an exponential improvement in the efficiency of the noise approximation. Alternatively, we can choose the eigenvalues of $\Gamma$
freely and obtain a valid matrix $V$ with numerical optimization, which may result in even more faithful approximation of $1/f^\alpha$ noise.

\begin{figure}[tb]
\begin{picture}(220,160)
\put(0,10){\includegraphics[width=220pt]{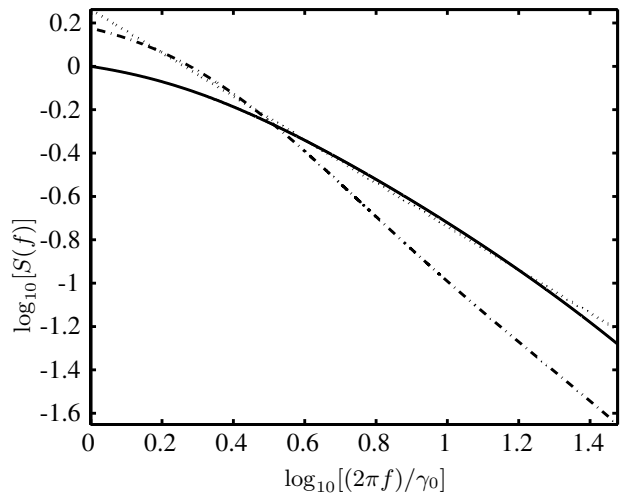}}
\put(93,0){$\log_{10}[(2\pi f)/\gamma_0]$} \put(-10,58){\rotatebox{90}{$\log_{10}[S(f)]$}}
\end{picture}
\caption{ Logarithm of the power spectral density for five independent RTN fluctuators
  (dash-dotted line), a multi-state Markovian source corresponding
  to~31 RTN fluctuators (solid line), and an ideal $1/f$ noise (dotted
  line). The transition rates of the RTN fluctuators are in both cases distributed
  uniformly on the interval $[\gamma_0,30\gamma_0]$.\label{fig:psd1}}
\label{fig:1}
\end{figure}

Figure~\ref{fig:1} compares the approximation of the spectral density of $1/f$ noise generated by independent RTN sources and by a multi-state
Markovian source. For the RTN approach, we choose 5 independent noise sources, for which the transition rates $\gamma_k$ are uniformly
distributed in the range $[\gamma_{\min}, \gamma_{\max}]=[\gamma_0, 30\gamma_0]$, and the strengths are given by $\Delta_k=1/\sqrt{\gamma_k}$.
This yields a fluctuator with 32 distinct noise states. For the multi-state fluctuator, we choose a 32-state noise source, for which the nonzero
eigenvalues $\lambda_k$ of its transition rate matrix $\Gamma$ are distributed uniformly on $[-60\gamma_0, -2\gamma_0]$, and $\chi_k =
1/\sqrt{-\lambda_k/2}$. Thus the condition in Eq.~\eqref{eq:1} is satisfied for both of the approaches and the multi-state noise source has an
autocorrelation function and power spectral density which are equal to those for a certain ensemble of 31 RTN fluctuators. We employ
representations of similar computational complexity here in order to be able to assess the relative accuracy for a given computational effort.

Figure~\ref{fig:1} shows that an ensemble of five RTN processes is not an accurate model for $1/f$ noise, whereas a single 32-state Markovian
noise source is quite accurate, especially in the range $3 \gamma_0 \lesssim \omega \lesssim 16 \gamma_0$.  The poor quality of the
approximation with five RTN fluctuators is due to the small number of independent noise sources employed here, whereas the 32-state Markovian
fluctuator contains more parameters and thereby introduces more flexibility in the noise approximation. The frequency range over which the
approximation is accurate is relatively short if one considers that the $1/f$ noise detected in experimental applications often extends over
several frequency decades. The width of this frequency range can of course be increased by increasing the width of the region from which the
eigenvalues of the matrix~$\Gamma$ are chosen. In this case, however, the number of discrete levels in the Markovian source must also be
increased to preserve the desired accuracy. For the main purpose of demonstrating the feasibility of the numerical optimization algorithm, in
the rest of this paper we will continue to approximate $1/f^\alpha$ noise by a single Markovian noise source with~32 levels.

\section{Qubit dynamics and control}
\label{sc:dyna}


In Ref.~\cite{hifid}, the temporal evolution of the qubit density
matrix was calculated by averaging over $10^4$--$10^5$ unitary quantum
trajectories, each corresponding to a sample noise path. To
ensure accuracy, a large number of unitary trajectories are required,
which results in extensive computational effort. In
Ref.~\cite{equdyn}, exact deterministic master equations describing
the average temporal evolution of quantum systems under Markovian
noise were derived.

Following Ref.~\cite{equdyn}, we introduce a conditional density
operator $\rho_k(t)$ which corresponds to the density operator of the
system averaged over all the noise sample paths occupying the $k$th
state at the time instant~$t$. The conditional density operators are
normalized such that the trace of the operator $\rho_{k}(t)$ yields
the probability of the $k$th noise state as $P_k(t) =
\tracep{\rho_k(t)}$.  The total average density operator can be
expressed as
\begin{equation}
  \label{eq:5}
\rho(t) = \sum_k \rho_k(t).
\end{equation}
The dynamics of $\rho_k$ is obtained from the coupled master
equations~\cite{equdyn}
\begin{equation}
\label{eq:markoveq}
\partial_t \rho_k(t) =
\frac{1}{i \hbar} \comm{H_k(t)}{\rho_k(t)} +\sum_{j} \Gamma_{kj} \rho_j(t),
\end{equation}
where $H_k(t)$ is the Hamiltonian of the system corresponding to the
$k$th noise state, and $\Gamma_{kj}$ the transition rate from the
$j$th state to the $k$th state, as defined in Sec.~\ref{sc:model}.
Specifically, in our one-qubit case,
\begin{equation}
  \label{eq:4}
  H_k(t) =\half a(t)\sigma_x+\half b_k \sigma_z,
\end{equation}
where $b_k$ is the noise amplitude of the state~$k$.
We shall use $\evol[a]{\rho}$ to denote the state $\rho$ evolved under
the influence of noise and the control sequence $a$.

The fidelity function quantifying the overlap between the desired
state $\rhof$ and the actual achieved final state is defined as
\begin{equation}
\label{eq:avefid}
\phi(\rhof,\evol[a]{\rho_0}) = \tracep{\rhof^\dagger \evol[a]{\rho_0}},
\end{equation}
where $\rho_0$ is the initial state of the system. To measure how
close the evolution $\mathcal{E}_a$ is to the intended quantum gate
operation $\Uf$, we calculate the average of the fidelity $\phi(\Uf
\rho_0 \Uf^\dagger, \evol[a]{\rho_0})$ over all pure initial
states~$\rho_0$, and obtain the gate fidelity function~\cite{hifid}
\begin{equation}
\label{eq:avegatefid} \Phi(\Uf)=\frac{1}{2}+\frac{1}{12} \sum_{k=x,y,z}\tracep{\Uf \sigma_k \Uf^\dagger \evol[a]{\sigma_k}}.
\end{equation}

We aim to find the optimal control pulses which maximize the fidelity of the achieved quantum operation, and hence apply a typical gradient
based optimization algorithm such as the gradient ascent pulse engineering (GRAPE) method developed in Ref.~\cite{grape}. If the continuous
pulse profiles are approximated by piecewise constant functions, the gradient of the fidelity function with respect to these constant pulse
values and durations can be calculated by the chain rule. This gradient is further used as a proportional adjustment to update the control pulse
profile. The optimization procedure is terminated when certain desired accuracy is achieved.
Note that due to the non-convex nature of the problem, the gradient based algorithm will only yield a locally optimal solution. We further
employ a multitude of initial conditions to find a control pulse which achieves the highest fidelity.

\section{Quantum memory}
\label{sc:results}

In this section, we focus on the implementation of quantum memory,
\ie, the identity operator. For the purpose of comparison with the
optimized pulse sequences, we introduce four other kinds of control
schemes which generate the identity operator.

The first reference sequence is simply not to apply any external control pulse, \ie, $a(t) = 0$. This pulse has no compensation for decoherence
or error. The second reference sequence is a constant $2\pi$ pulse given by
\begin{equation}
\label{eq:2pi-pulse}
a_{2\pi}(t)=a_{\max},\quad \text{for } t \in [0, 2\pi\hbar/a_{\max}].
\end{equation}
The third reference sequence is the composite pulse sequence known as compensation for off-resonance with a pulse sequence (CORPSE), which was
originally designed to correct systematic errors in the implementation of one-qubit quantum operations and to provide high order control
protocols for systematic qubit bias, \ie, for the noise correlation time $\tau_c \rightarrow \infty$~\cite{cummins2000,
  cummins2003}. For the identity operation, the CORPSE pulse sequence
can be obtained as
\begin{equation}\label{eq:2picorpse}
a_{\rm SC2\pi}(t)=\left\{
\begin{array}{rll}
a_{\rm max}, & {\rm for} & 0<t'< \pi \\
-a_{\rm max}, & {\rm for} & \pi \le t'\le 3\pi \\
a_{\rm max}, & {\rm for} & 3\pi<t'< 4\pi,
\end{array}\right.
\end{equation}
where the dimensionless time $t'$ is defined as $t'=a_{\max} t/\hbar$.

In the absence of noise, the CORPSE sequence generates the identity operator exactly although it requires twice as long operation time as a
$2\pi$ pulse, the second reference pulse above. In the presence of small systematic errors, the CORPSE sequence is much more accurate than the
$2\pi$ pulse. For example, consider a state transformation from the north pole back to itself on the Bloch sphere. For $\eta(t)\equiv\Delta$ in
Eq.~\eqref{eq:qubham}, the fidelities defined in Eq.~\eqref{eq:avefid} can be derived to be
\begin{eqnarray}
  \label{eq:3}
 \phi_{2\pi}=1-\frac{\pi^2}{4}\left(\frac{\Delta}{a_{\max}} \right)^4
+O\left(\frac{\Delta}{a_{\max}} \right)^6,
\end{eqnarray}
and
\begin{equation}
  \label{eq:6}
   \phi_{\textrm{SC}2\pi}=1-4\pi^2\left(\frac{\Delta}{a_{\max}} \right)^8
+O\left(\frac{\Delta}{a_{\max}} \right)^{10}.
\end{equation}
We observe that the error in the fidelity of the $2\pi$ pulse is fourth order in the relative noise strength $\Delta/a_{\rm max}$, whereas for
the CORPSE pulse sequence it is eighth order. Thus the CORPSE sequence is much more accurate than a $2\pi$ pulse in correcting the effects of
systematic errors on quantum memory.

The fourth standard pulse sequence which we take as a reference is the Carr-Purcell-Meiboom-Gill (CPMG)~\cite{meiboom58} sequence which is
designed to preserve qubit coherence. In our context, this sequence consists of a $\pi/2$ pulse followed by multiple $\pi$ pulses at intervals
$t_p$, followed by a final $\pi/2$ pulse to bring the system back to the original state.  This pulse sequence is designed for $T_2$ measurements
on spins, starting from the $|0\rangle$ state. Thus one does not expect a CPMG pulse sequence to perform as well if the initial state is
averaged over the Bloch sphere as is done to compute a gate fidelity.

\begin{figure}[tb]
\psfrag{X}[][]{$\tau_{\rm c}/(\hbar/a_{\rm max})$}
\psfrag{Y}[][]{$\Phi(I)$}
\includegraphics[width=220pt]{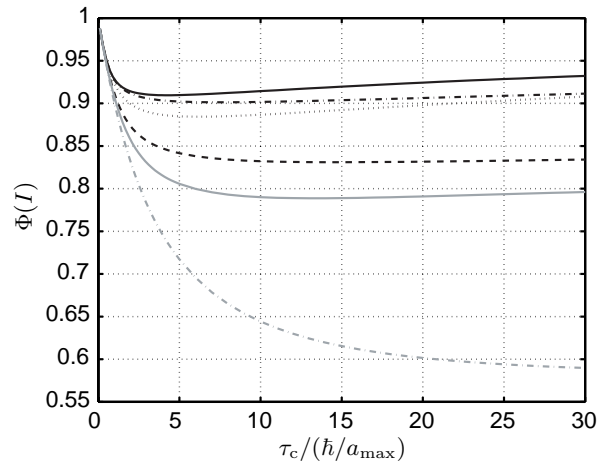}
\caption{Fidelity of the quantum memory as a function of the
  characteristic correlation time $\tau_{\rm c}$ for optimized control
  pulses (black solid), a $2\pi$ pulse (black dash-dotted), CORPSE
  pulse sequence (black dotted), CPMG1 pulse sequence (black dashed),
  CPMG2 pulse sequence (gray solid), zero pulse sequence (gray
  dash-dotted). The operation time is chosen to be
  $12\pi\hbar/a_\textrm{max}$. The noise is produced by a single
  32-state Markovian source with the average strength
  $\ev{|\eta|}=0.125\times a_{\rm max}$ corresponding to 31 RTN
  fluctuators with the transition rates uniformly distributed over the
  region $[1/\tau_\textrm{c},30/\tau_\textrm{c}]$ and strengths chosen
  as described in Sec.~\ref{sc:model}.}
\label{fig:2}
\end{figure}

We first present the fidelities obtained for the identity operator using the various control pulse options in the presence of $1/f$ noise. The
noise is generated here by the single Markovian noise source discussed in Sec.~\ref{sc:model}, with transition rates distributed uniformly over
the interval $[1/\tau_c, 30/\tau_c]$ .  In Fig.~\ref{fig:2}, the fidelities obtained from optimized control pulses, $2\pi$ pulse, CORPSE, CPMG,
and zero pulse sequences are plotted as functions of the characteristic correlation time $\tau_c$ of the approximate $1/f$ noise. Here, CPMG1
and CPMG2 refer to two CPMG types of pulses with the intervals between $\pi$ pulses being $\pi$ and $2\pi$, respectively.  The total duration
for these pulses are all $12\pi\hbar/a_\textrm{max}$. The optimal control pulse is designed for $6\pi$, and therefore we repeat it twice.
Similarly, we repeat the $2\pi$ pulse 6 times, the CORPSE sequence 3 times, the CPMG1 sequence 3 times, and the CPMG2 sequence twice. The
optimal control pulse yields clearly the highest fidelity among all these pulses, whereas the zero pulse sequence has the worst performance as
there are no correction mechanisms.  Note that due to motional narrowing, all curves approach unit fidelity in the limit $\tau_{\rm c} \to 0$.

\begin{figure}[tb]
\centering
  \psfrag{X}[][]{$T/(\pi\hbar/a_{\rm max})$}
  \psfrag{Y}[][]{$\Phi(I)$}
\includegraphics[width=220pt]{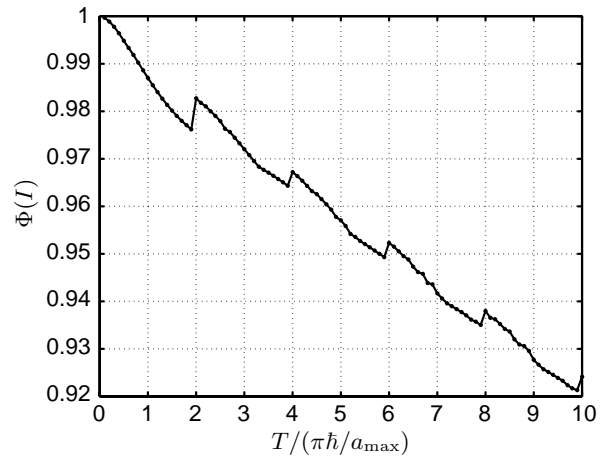}
\caption{Fidelity of the quantum memory as a function of the operation
  time for control pulses optimized at each point. The noise is
  produced by a similar multi-state Markovian source as in
  Fig.~\ref{fig:2}, with $\tau_c=3\hbar/a_{\max}$.
 \label{fig:3}}
\end{figure}

The memory access rate is an important specification in modern computer technology~\cite{Hennessy06}. In our context, it corresponds to the
total duration of the control pulses.  Figure~\ref{fig:3} shows the fidelity as a function of the duration for the numerically optimized control
pulses.  Equation~\eqref{eq:qubham} implies that in the absence of noise, the quantum system will generate an identity operator for $a=a_{\rm
max}$ and the duration $T=2n\pi/a_{\rm max}$. In Fig.~\ref{fig:3}, we observe that, despite an overall decrease, there are peaks in the fidelity
near these operation times.  Thus we can regard $2n\pi/a_{\rm max}$ as the natural periods for quantum memory, and we always choose the total
duration of control pulses correspondingly.

\begin{figure}[tb]
\centering \psfrag{X}[][]{$\ev{|\eta|}$} \psfrag{Y}[][]{$\Phi(I)$}
\includegraphics[width=220pt]{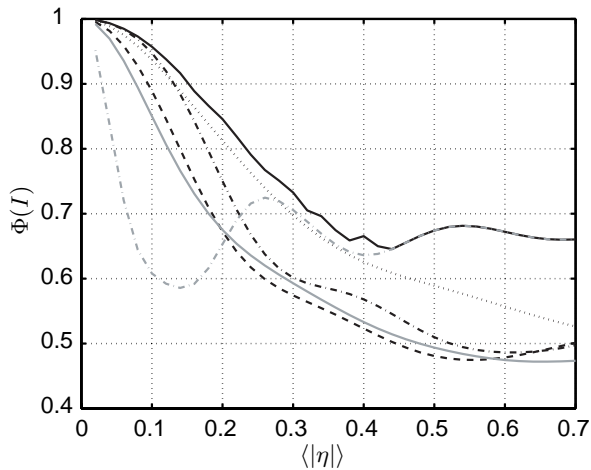}
\caption{Fidelity of the quantum memory as a function of the average
  absolute noise strength for optimized control pulses (black solid),
  $2\pi$ pulse (black dash-dotted), CORPSE (black dotted), CPMG1 (black
  dashed), CPMG2 (gray solid), and zero (gray dash-dotted). The
  operation time is chosen to be $12\pi\hbar/a_\textrm{max}$. Except
  for its strength, the noise is produced by a similar multi-state
  Markovian source as in Fig.~\ref{fig:2} with the correlation time
  $\tau_\textrm{c} = 30\hbar/a_\textrm{max}$.}
\label{fig:5}
\end{figure}

Here, we study the relation between the optimized fidelities achieved above and the average noise strength $\ev{|\eta|}$, for a fixed value of
the characteristic correlation time $\tau_{\rm c}= 30 \hbar/a_{\rm max}$. Figure~\ref{fig:5} shows the fidelity as a function of the noise
strength for the optimized control pulses, $2\pi$ pulse, the CORPSE, CPMG1, CPMG2, and zero pulse sequences. At small values of $\ev{|\eta|}$
again, the optimized control pulses consistently achieve higher fidelities than all reference pulses. However, we note that if the noise
strength exceeds $\sim$0.4, the optimized pulse sequence reduces to the zero pulse sequence, i.e., any nonzero pulse sequence will actually
deteriorate the fidelity performance.

\begin{figure}[tb]
  \centering \psfrag{X}[][]{$\tau_{\rm c}/(\hbar/a_{\rm max})$}
  \psfrag{Y}[][]{$\Phi(I)$} \includegraphics[width=220pt]{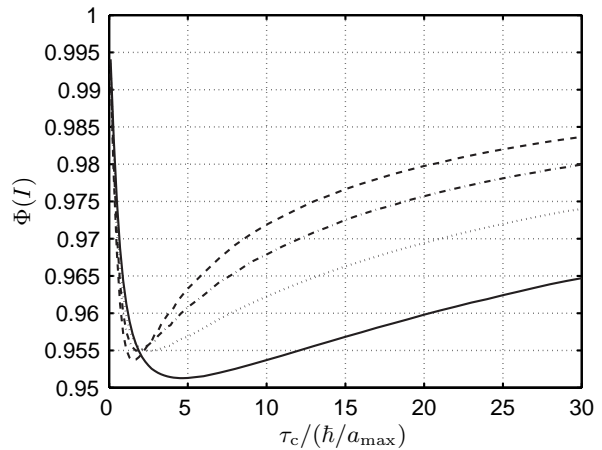}
\caption{Fidelity of the quantum memory achieved with optimized
  control pulses as a function of the characteristic correlation time
  $\tau_{\rm c}$ for $1/f^\alpha$ noise with $\alpha=1$ (solid
 ), $\alpha=1.25$ (dotted), $\alpha=1.5$ (dash-dotted),
  and $\alpha=1.75$ (dashed). The operation time is chosen to be
  $6\pi\hbar/a_\textrm{max}$. The noise is produced by a similar multi-state Markovian source as
  in Fig.~\ref{fig:2} with variable values of the power~$\alpha$.} \label{fig:4}
\end{figure}

The discussion above is based on the specific noise density spectrum $1/f^\alpha$ with $\alpha=1$.  Figure~\ref{fig:4} shows the fidelities of
quantum memory for four optimized control pulses, each of which is obtained for a different value of $\alpha$. The noise is produced here by a
single multi-state Markovian source with average strength $\ev{|\eta|} = 0.125\times a_{\rm max}$, and the total duration for all control pulses
are fixed to~$6\pi$. A systematic scaling of the correlation time axis with respect to~$\alpha$ is clearly visible in Fig.~\ref{fig:4}. This
phenomenon is explained by the fact that the concentration of the power spectrum of $1/f^\alpha$ to high frequencies, \ie, long correlation
times, increases with~$\alpha$. Hence, the curves scale down in~$\tau_\textrm{c}$ with increasing~$\alpha$.


\section{{\sc NOT} gate}
\label{sc:resultsNOT} In this section, we focus on the generation of high-fidelity {\sc NOT} gates, \ie, the $\sigma_x$ operator, under $1/f$
noise. As in the case of quantum memory, we compare the numerically optimized results with reference pulses. In this case, our first reference
pulse is the $\pi$ pulse given by \be \label{eq:pi-pulse} a_\pi(t)=a_{\rm max},\quad {\rm
  for}\quad t\in[0, \pi\hbar/a_{\rm max}], \ee which in the absence of
noise is the most efficient way of achieving a {\sc NOT} gate. In addition, we will use the two composite pulse sequences CORPSE and short
CORPSE \cite{cummins2000,cummins2003} which assume here the form \be\label{eq:corpse} a_{\rm C\pi}(t)=\left\{
\begin{array}{rll}
a_{\rm max}, & {\rm for} & 0<t'< \pi/3 \\
-a_{\rm max}, & {\rm for} & \pi/3\le t'\le 2\pi \\
a_{\rm max}, & {\rm for} & 2\pi<t'< 13\pi/3,
\end{array}\right.
\ee and \be\label{eq:scorpse} a_{\rm SC\pi}(t)=\left\{
\begin{array}{rll}
-a_{\rm max}, & {\rm for} & 0<t'< \pi/3 \\
a_{\rm max}, & {\rm for} & \pi/3\le t'\le 2\pi \\
-a_{\rm max}, & {\rm for} & 2\pi<t'< 7\pi/3,
\end{array}\right.
\ee respectively. Both of these pulse sequences correct for systematic error, CORPSE being more efficient. However, the operation time of short
CORPSE is much shorter than that of CORPSE, and hence it can yield higher fidelities in the presence of noise.

\begin{figure}[tbh]
\begin{picture}(220,180)
\put(0,10){\includegraphics[width=220pt]{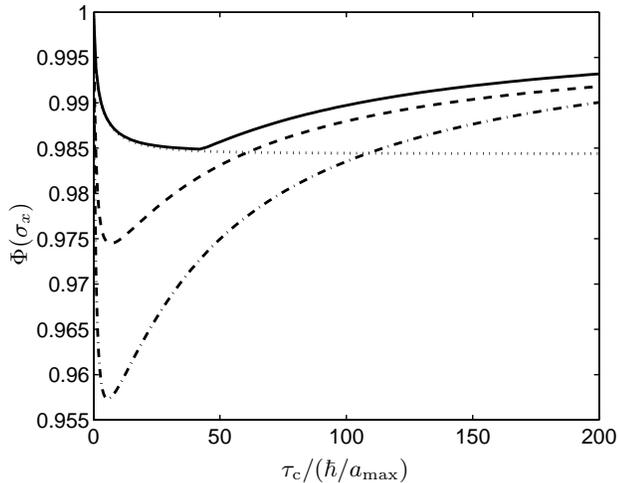}}
\put(93,0){$\tau_{\rm c}/(\hbar/a_{\rm max})$}
\put(-10,80){\rotatebox{90}{$\Phi(\sigma_x)$}}
\end{picture}
\caption{\label{fig:6} {\sc NOT} gate fidelities as
  functions of the characteristic noise correlation time $\tau_{\rm
    c}$ for a $\pi$ pulse (dotted), CORPSE (dash-dotted), short CORPSE (dashed),
  and gradient optimized pulse sequence (solid). The $1/f$ noise is
  generated as in Fig.~\ref{fig:2}.}
\end{figure}

\begin{figure}[tbh]
\includegraphics[width=220pt]{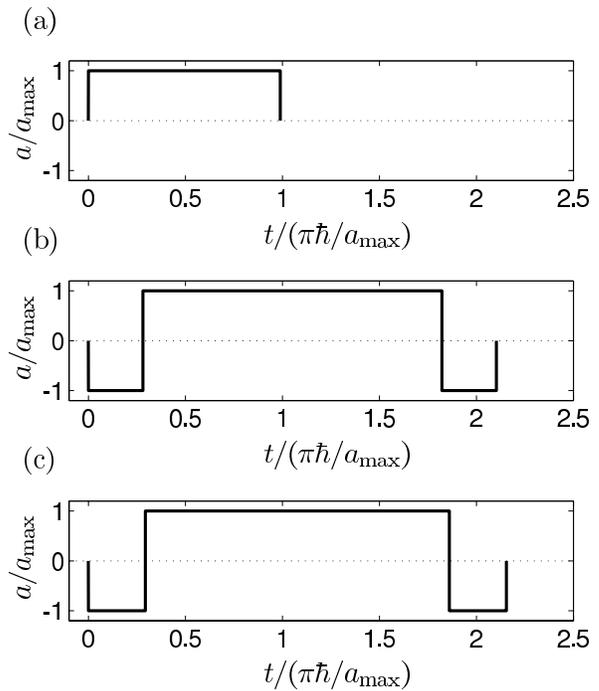}
\caption{\label{fig:7} Optimized pulse sequences yielding the highest
gate fidelities for correlation times (a) $45\hbar/a_\textrm{max}$,
(b) $100\hbar/a_\textrm{max}$, and (c) $150\hbar/a_\textrm{max}$
corresponding to Fig.~\ref{fig:6}.}
\end{figure}

Figure~\ref{fig:6} shows the {\sc NOT} gate fidelities obtained by the reference and optimized pulses in the presence of the same~$1/f$ noise as
employed in the analysis of quantum memory in Sec.~\ref{sc:results}. We observe that for long enough correlation times, the composite pulse
sequences provide good error correction. Furthermore, as observed earlier for RTN~\cite{hifid}, for intermediate correlation times, short CORPSE
achieves the highest fidelity among the reference pulses. Figure~\ref{fig:7} presents the pulse sequences obtained from the numerical
optimizations for three different values of the noise correlation time $\tau_c$.  For the optimized pulse sequence, we find a transition from an
approximately constant pulse to a short CORPSE -like pulse sequence at characteristic correlation time $\tau_\textrm{c}\approx
50\hbar/a_\textrm{max}$. This change in optimal pulse sequence is responsible for the apparent discontinuity in the first derivative of the
fidelity curve in Fig.~\ref{fig:6}.

These results for the generation of {\sc NOT} gates under $1/f$ noise are qualitatively quite similar to the previous results presented in
Refs.~\cite{hifid,equdyn} for a single RTN. This similarity is due to the fact that $1/f$ noise can be regarded as arising from a sum of
independent RTN fluctuators, each of which having a similar fidelity dependence on their correlation times.  Note that the scale for the
reference correlation time $\tau_c$ of the fidelity obtained in presence of $1/f$ noise in Fig.~\ref{fig:6} is somewhat different from the
corresponding scale for the correlation time of a single RTN source, since the $1/f$ noise involves an ensemble of RTN fluctuators with a range
of correlation times.

\section{Conclusions}
\label{sc:discussion}

We have studied a single qubit under the influence of $1/f^\alpha$ noise for $2> \alpha > 0$ and investigated how decoherence due to this noise
can be suppressed in the implementation of single qubit operations. We presented an efficient way to approximate the noise with a discrete
multi-state Markovian fluctuator.  Due to this finding, the average temporal evolution of the qubit density matrix under $1/f^\alpha$ noise can
be efficiently determined from a deterministic master equation.

Employing these exact deterministic master equations describing the temporal evolution of the qubit density operator under Markovian noise, we
applied a gradient based optimization procedure to search for optimal control pulses implementing quantum operations. In particular, we studied
the physical application of quantum memory, \ie, the identity operator, which is a fundamental concept in the realization of a quantum computer.
The optimized control pulses significantly improved the fidelity over several reference sequences such as $2\pi$, CORPSE, CPMG, and zero pulses.
We observe peaks on fidelity curves corresponding to integer multiples of $2\pi\hbar/a_\textrm{max}$ in the total durations of control pulses,
where $a_\textrm{max}$ is the maximum magnitude of the external control field. We also studied the performance of optimal control pulses under
$1/f^\alpha$ noise for several different values of $2>\alpha \geq 1$, and found a monotonic behavior in the noise frequency as a function of
$\alpha$, \ie, the fidelity curves are scaled down in the correlation time for increasing $\alpha$. We also investigated how the fidelities
degraded as the noise strength increases. For the generation of high-fidelity {\sc NOT} gates, we obtained results showing qualitatively similar
behavior to the previous results presented in Refs.~\cite{hifid,equdyn} for a single RTN source. In particular, just as for a single noise
source, in the presence of $1/f^\alpha$ noise we observed a transition in the optimal control pulse sequence from a constant pulse to a
CORPSE-like sequence as the noise characteristic correlation time $\tau_c$ is increased.

This approach of coupled master equations indexed by noise states of the environment, together with an optimization technique for pulse design
can be readily generalized to multiple qubits evolving in the presence of $1/f^\alpha$ noise and other Markovian noise sources. Furthermore, it
can be used to develop realistic pulse sequences for mitigation of nuclear spin and surface magnetic noise acting on donor spins implanted in
silicon~\cite{Schenkel06}, as well as for suppression of background charge noise acting on superconducting qubits~\cite{astafiev06}. In future,
we will study the implementation of multi-qubit gates, \eg, the controlled {\sc NOT} gate, in noisy systems and the swapping of quantum
information from a noisy qubit to long term quantum memory. We will also consider more realistic noise with $1/f^\alpha$ spectrum over many
frequency decades.

\begin{acknowledgements}
This work was supported by the Academy of Finland, the National Security Agency (NSA) under MOD713106A and by the NSF ITR program under grant
number EIA-0205641. M.\ M.\ and V.\ B.\ acknowledge the Finnish Cultural Foundation, M.\ M.\ the V\"ais\"al\"a foundation and Magnus Ehrnrooth
Foundation for the financial support. We thank J.\ Clarke for insightful discussions.
\end{acknowledgements}

\bibliographystyle{apsrev}
\bibliography{refs}

\begin{thebibliography}{35}
\expandafter\ifx\csname natexlab\endcsname\relax\def\natexlab#1{#1}\fi
\expandafter\ifx\csname bibnamefont\endcsname\relax
  \def\bibnamefont#1{#1}\fi
\expandafter\ifx\csname bibfnamefont\endcsname\relax
  \def\bibfnamefont#1{#1}\fi
\expandafter\ifx\csname citenamefont\endcsname\relax
  \def\citenamefont#1{#1}\fi
\expandafter\ifx\csname url\endcsname\relax
  \def\url#1{\texttt{#1}}\fi
\expandafter\ifx\csname urlprefix\endcsname\relax\def\urlprefix{URL }\fi
\providecommand{\bibinfo}[2]{#2}
\providecommand{\eprint}[2][]{\url{#2}}

\bibitem[{\citenamefont{Faoro and Viola}(2004)}]{faoroviola}
\bibinfo{author}{\bibfnamefont{L.}~\bibnamefont{Faoro}} \bibnamefont{and}
  \bibinfo{author}{\bibfnamefont{L.}~\bibnamefont{Viola}},
  \bibinfo{journal}{Phys. Rev. Lett.} \textbf{\bibinfo{volume}{92}},
  \bibinfo{pages}{117905} (\bibinfo{year}{2004}).

\bibitem[{\citenamefont{Paladino
  et~al.}(2002{\natexlab{a}})\citenamefont{Paladino, Faoro, Falci, and
  Fazio}}]{Paladino2002}
\bibinfo{author}{\bibfnamefont{E.}~\bibnamefont{Paladino}},
  \bibinfo{author}{\bibfnamefont{L.}~\bibnamefont{Faoro}},
  \bibinfo{author}{\bibfnamefont{G.}~\bibnamefont{Falci}}, \bibnamefont{and}
  \bibinfo{author}{\bibfnamefont{R.}~\bibnamefont{Fazio}},
  \bibinfo{journal}{Phys. Rev. Lett.} \textbf{\bibinfo{volume}{88}},
  \bibinfo{pages}{228304} (\bibinfo{year}{2002}{\natexlab{a}}).

\bibitem[{\citenamefont{Falci et~al.}(2004)\citenamefont{Falci,
  D\char39{}Arrigo, Mastellone, and Paladino}}]{Falci2004}
\bibinfo{author}{\bibfnamefont{G.}~\bibnamefont{Falci}},
  \bibinfo{author}{\bibfnamefont{A.}~\bibnamefont{D\char39{}Arrigo}},
  \bibinfo{author}{\bibfnamefont{A.}~\bibnamefont{Mastellone}},
  \bibnamefont{and} \bibinfo{author}{\bibfnamefont{E.}~\bibnamefont{Paladino}},
  \bibinfo{journal}{Phys. Rev. A} \textbf{\bibinfo{volume}{70}},
  \bibinfo{pages}{040101} (\bibinfo{year}{2004}).

\bibitem[{\citenamefont{Astafiev et~al.}(2006)\citenamefont{Astafiev, Pashkin,
  Nakamura, Yamamoto, and Tsai}}]{astafiev06}
\bibinfo{author}{\bibfnamefont{O.}~\bibnamefont{Astafiev}},
  \bibinfo{author}{\bibfnamefont{Y.~A.} \bibnamefont{Pashkin}},
  \bibinfo{author}{\bibfnamefont{Y.}~\bibnamefont{Nakamura}},
  \bibinfo{author}{\bibfnamefont{T.}~\bibnamefont{Yamamoto}}, \bibnamefont{and}
  \bibinfo{author}{\bibfnamefont{J.~S.} \bibnamefont{Tsai}},
  \bibinfo{journal}{Phys. Rev. Lett.} \textbf{\bibinfo{volume}{96}},
  \bibinfo{pages}{137001} (\bibinfo{year}{2006}).

\bibitem[{\citenamefont{Astafiev et~al.}(2004)\citenamefont{Astafiev, Pashkin,
  Nakamura, Yamamoto, and Tsai}}]{astafiev04}
\bibinfo{author}{\bibfnamefont{O.}~\bibnamefont{Astafiev}},
  \bibinfo{author}{\bibfnamefont{Y.~A.} \bibnamefont{Pashkin}},
  \bibinfo{author}{\bibfnamefont{Y.}~\bibnamefont{Nakamura}},
  \bibinfo{author}{\bibfnamefont{T.}~\bibnamefont{Yamamoto}}, \bibnamefont{and}
  \bibinfo{author}{\bibfnamefont{J.~S.} \bibnamefont{Tsai}},
  \bibinfo{journal}{Phys. Rev. Lett.} \textbf{\bibinfo{volume}{93}},
  \bibinfo{pages}{267007} (\bibinfo{year}{2004}).

\bibitem[{\citenamefont{Wellstood et~al.}(2004)\citenamefont{Wellstood, Urbina,
  and Clarke}}]{wellstood04}
\bibinfo{author}{\bibfnamefont{F.~C.} \bibnamefont{Wellstood}},
  \bibinfo{author}{\bibfnamefont{C.}~\bibnamefont{Urbina}}, \bibnamefont{and}
  \bibinfo{author}{\bibfnamefont{J.}~\bibnamefont{Clarke}},
  \bibinfo{journal}{Apl. Phys. Lett.} \textbf{\bibinfo{volume}{85}},
  \bibinfo{pages}{5296} (\bibinfo{year}{2004}).

\bibitem[{\citenamefont{M\"{u}ck et~al.}(2005)\citenamefont{M\"{u}ck, Korn,
  Mugford, Kycia, and Clarke}}]{muck05}
\bibinfo{author}{\bibfnamefont{M.}~\bibnamefont{M\"{u}ck}},
  \bibinfo{author}{\bibfnamefont{M.}~\bibnamefont{Korn}},
  \bibinfo{author}{\bibfnamefont{C.~G.~A.} \bibnamefont{Mugford}},
  \bibinfo{author}{\bibfnamefont{J.~B.} \bibnamefont{Kycia}}, \bibnamefont{and}
  \bibinfo{author}{\bibfnamefont{J.}~\bibnamefont{Clarke}},
  \bibinfo{journal}{Apl. Phys. Lett.} \textbf{\bibinfo{volume}{86}},
  \bibinfo{pages}{012610} (\bibinfo{year}{2005}).

\bibitem[{\citenamefont{Eiles et~al.}(1992)\citenamefont{Eiles, Kautz, and
  Martinis}}]{zimmerli92}
\bibinfo{author}{\bibfnamefont{T.~M.} \bibnamefont{Eiles}},
  \bibinfo{author}{\bibfnamefont{R.~L.} \bibnamefont{Kautz}}, \bibnamefont{and}
  \bibinfo{author}{\bibfnamefont{J.~M.} \bibnamefont{Martinis}},
  \bibinfo{journal}{Apl. Phys. Lett.} \textbf{\bibinfo{volume}{61}},
  \bibinfo{pages}{237} (\bibinfo{year}{1992}).

\bibitem[{\citenamefont{Nakamura
  et~al.}(2002{\natexlab{a}})\citenamefont{Nakamura, Pashkin, Yamamoto, and
  Tsai}}]{nakamura}
\bibinfo{author}{\bibfnamefont{Y.}~\bibnamefont{Nakamura}},
  \bibinfo{author}{\bibfnamefont{Y.~A.} \bibnamefont{Pashkin}},
  \bibinfo{author}{\bibfnamefont{T.}~\bibnamefont{Yamamoto}}, \bibnamefont{and}
  \bibinfo{author}{\bibfnamefont{J.~S.} \bibnamefont{Tsai}},
  \bibinfo{journal}{Physica Scripta} \textbf{\bibinfo{volume}{102}},
  \bibinfo{pages}{155} (\bibinfo{year}{2002}{\natexlab{a}}).

\bibitem[{\citenamefont{de~Sousa}(2006)}]{deSousa06}
\bibinfo{author}{\bibfnamefont{R.}~\bibnamefont{de~Sousa}},
  \bibinfo{journal}{unpublished, cond-mat/0610716}  (\bibinfo{year}{2006}).

\bibitem[{\citenamefont{Schenkel et~al.}(2006)\citenamefont{Schenkel, Liddle,
  Persaud, Tyryshkin, Lyon, de~Sousa, Whaley, Shangkuan, and
  Chakarov}}]{Schenkel06}
\bibinfo{author}{\bibfnamefont{T.}~\bibnamefont{Schenkel}},
  \bibinfo{author}{\bibfnamefont{J.~A.} \bibnamefont{Liddle}},
  \bibinfo{author}{\bibfnamefont{A.}~\bibnamefont{Persaud}},
  \bibinfo{author}{\bibfnamefont{A.~M.} \bibnamefont{Tyryshkin}},
  \bibinfo{author}{\bibfnamefont{S.~A.} \bibnamefont{Lyon}},
  \bibinfo{author}{\bibfnamefont{R.}~\bibnamefont{de~Sousa}},
  \bibinfo{author}{\bibfnamefont{K.~B.} \bibnamefont{Whaley}},
  \bibinfo{author}{\bibfnamefont{J.~B.~J.} \bibnamefont{Shangkuan}},
  \bibnamefont{and} \bibinfo{author}{\bibfnamefont{I.}~\bibnamefont{Chakarov}},
  \bibinfo{journal}{Apl. Phys. Lett.} \textbf{\bibinfo{volume}{8}},
  \bibinfo{pages}{11201} (\bibinfo{year}{2006}).

\bibitem[{\citenamefont{de~Sousa~{\it et al.}}(2007)}]{deSousa07}
\bibinfo{author}{\bibfnamefont{R.}~\bibnamefont{de~Sousa~{\it et al.}}},
  \bibinfo{journal}{unpublished}  (\bibinfo{year}{2007}).

\bibitem[{\citenamefont{Nakamura
  et~al.}(2002{\natexlab{b}})\citenamefont{Nakamura, Pashkin, Yamamoto, and
  Tsai}}]{Nakamura2002}
\bibinfo{author}{\bibfnamefont{Y.}~\bibnamefont{Nakamura}},
  \bibinfo{author}{\bibfnamefont{Y.~A.} \bibnamefont{Pashkin}},
  \bibinfo{author}{\bibfnamefont{T.}~\bibnamefont{Yamamoto}}, \bibnamefont{and}
  \bibinfo{author}{\bibfnamefont{J.~S.} \bibnamefont{Tsai}},
  \bibinfo{journal}{Phys. Rev. Lett.} \textbf{\bibinfo{volume}{88}},
  \bibinfo{pages}{047901} (\bibinfo{year}{2002}{\natexlab{b}}).

\bibitem[{\citenamefont{Galperin
  et~al.}(2006{\natexlab{a}})\citenamefont{Galperin, Altshuler, Bergli, and
  Shantsev}}]{Galperin03}
\bibinfo{author}{\bibfnamefont{Y.~M.} \bibnamefont{Galperin}},
  \bibinfo{author}{\bibfnamefont{B.~L.} \bibnamefont{Altshuler}},
  \bibinfo{author}{\bibfnamefont{J.}~\bibnamefont{Bergli}}, \bibnamefont{and}
  \bibinfo{author}{\bibfnamefont{D.~V.} \bibnamefont{Shantsev}},
  \bibinfo{journal}{Phys. Rev. Lett.} \textbf{\bibinfo{volume}{96}},
  \bibinfo{pages}{097009} (\bibinfo{year}{2006}{\natexlab{a}}).

\bibitem[{\citenamefont{Savo et~al.}(1987)\citenamefont{Savo, Wellstood, and
  Clarke}}]{Savo1987}
\bibinfo{author}{\bibfnamefont{B.}~\bibnamefont{Savo}},
  \bibinfo{author}{\bibfnamefont{F.~C.} \bibnamefont{Wellstood}},
  \bibnamefont{and} \bibinfo{author}{\bibfnamefont{J.}~\bibnamefont{Clarke}},
  \bibinfo{journal}{Appl. Phys. Letts.} \textbf{\bibinfo{volume}{50}},
  \bibinfo{pages}{1757} (\bibinfo{year}{1987}).

\bibitem[{\citenamefont{Wakai and Harlingen}(1987)}]{Wakai1987}
\bibinfo{author}{\bibfnamefont{R.~T.} \bibnamefont{Wakai}} \bibnamefont{and}
  \bibinfo{author}{\bibfnamefont{D.~J.~V.} \bibnamefont{Harlingen}},
  \bibinfo{journal}{Phys. Rev. Lett.} \textbf{\bibinfo{volume}{58}},
  \bibinfo{pages}{1687} (\bibinfo{year}{1987}).

\bibitem[{\citenamefont{Fujisawa and Hirayama}(2000)}]{Fujisawa2000}
\bibinfo{author}{\bibfnamefont{T.}~\bibnamefont{Fujisawa}} \bibnamefont{and}
  \bibinfo{author}{\bibfnamefont{Y.}~\bibnamefont{Hirayama}},
  \bibinfo{journal}{Appl. Phys. Lett.} \textbf{\bibinfo{volume}{77}},
  \bibinfo{pages}{543} (\bibinfo{year}{2000}).

\bibitem[{\citenamefont{Kurdak et~al.}(1997)\citenamefont{Kurdak, Chen, Tsui,
  Parihar, Lyon, and Weimann}}]{Kurdak1997}
\bibinfo{author}{\bibfnamefont{C.}~\bibnamefont{Kurdak}},
  \bibinfo{author}{\bibfnamefont{C.-J.} \bibnamefont{Chen}},
  \bibinfo{author}{\bibfnamefont{D.~C.} \bibnamefont{Tsui}},
  \bibinfo{author}{\bibfnamefont{S.}~\bibnamefont{Parihar}},
  \bibinfo{author}{\bibfnamefont{S.}~\bibnamefont{Lyon}}, \bibnamefont{and}
  \bibinfo{author}{\bibfnamefont{G.~W.} \bibnamefont{Weimann}},
  \bibinfo{journal}{Phys. Rev. Lett.} \textbf{\bibinfo{volume}{56}},
  \bibinfo{pages}{9813} (\bibinfo{year}{1997}).

\bibitem[{\citenamefont{de~Sousa et~al.}(2005)\citenamefont{de~Sousa, Whaley,
  Wilhelm, and von Delft}}]{deSousa05}
\bibinfo{author}{\bibfnamefont{R.}~\bibnamefont{de~Sousa}},
  \bibinfo{author}{\bibfnamefont{K.~B.} \bibnamefont{Whaley}},
  \bibinfo{author}{\bibfnamefont{F.~K.} \bibnamefont{Wilhelm}},
  \bibnamefont{and} \bibinfo{author}{\bibfnamefont{J.}~\bibnamefont{von
  Delft}}, \bibinfo{journal}{Phys. Rev. Lett.} \textbf{\bibinfo{volume}{95}},
  \bibinfo{pages}{247006} (\bibinfo{year}{2005}).

\bibitem[{\citenamefont{Viola and Lloyd}(1998)}]{violalloyd}
\bibinfo{author}{\bibfnamefont{L.}~\bibnamefont{Viola}} \bibnamefont{and}
  \bibinfo{author}{\bibfnamefont{S.}~\bibnamefont{Lloyd}},
  \bibinfo{journal}{Phys. Rev. A} \textbf{\bibinfo{volume}{58}},
  \bibinfo{pages}{2733} (\bibinfo{year}{1998}).

\bibitem[{\citenamefont{Viola et~al.}(1999)\citenamefont{Viola, Lloyd, and
  Knill}}]{violalloydknill}
\bibinfo{author}{\bibfnamefont{L.}~\bibnamefont{Viola}},
  \bibinfo{author}{\bibfnamefont{S.}~\bibnamefont{Lloyd}}, \bibnamefont{and}
  \bibinfo{author}{\bibfnamefont{E.}~\bibnamefont{Knill}},
  \bibinfo{journal}{Phys. Rev. Lett.} \textbf{\bibinfo{volume}{83}},
  \bibinfo{pages}{4888} (\bibinfo{year}{1999}).

\bibitem[{\citenamefont{Kofman and Kurizki}(2001)}]{Kofman01}
\bibinfo{author}{\bibfnamefont{A.~G.} \bibnamefont{Kofman}} \bibnamefont{and}
  \bibinfo{author}{\bibfnamefont{G.}~\bibnamefont{Kurizki}},
  \bibinfo{journal}{Phys. Rev. Lett.} \textbf{\bibinfo{volume}{87}},
  \bibinfo{pages}{270405} (\bibinfo{year}{2001}).

\bibitem[{\citenamefont{Kofman and Kurizki}(2004)}]{Kofman04}
\bibinfo{author}{\bibfnamefont{A.~G.} \bibnamefont{Kofman}} \bibnamefont{and}
  \bibinfo{author}{\bibfnamefont{G.}~\bibnamefont{Kurizki}},
  \bibinfo{journal}{Phys. Rev. Lett.} \textbf{\bibinfo{volume}{93}},
  \bibinfo{pages}{130406} (\bibinfo{year}{2004}).

\bibitem[{\citenamefont{M\"ott\"onen et~al.}(2006)\citenamefont{M\"ott\"onen,
  Sousa, Zhang, and Whaley}}]{hifid}
\bibinfo{author}{\bibfnamefont{M.}~\bibnamefont{M\"ott\"onen}},
  \bibinfo{author}{\bibfnamefont{R.~d.} \bibnamefont{Sousa}},
  \bibinfo{author}{\bibfnamefont{J.}~\bibnamefont{Zhang}}, \bibnamefont{and}
  \bibinfo{author}{\bibfnamefont{K.~B.} \bibnamefont{Whaley}},
  \bibinfo{journal}{Phys. Rev. A} \textbf{\bibinfo{volume}{73}},
  \bibinfo{pages}{022332} (\bibinfo{year}{2006}).

\bibitem[{\citenamefont{Weissman}(1988)}]{Weissman88}
\bibinfo{author}{\bibfnamefont{M.~B.} \bibnamefont{Weissman}},
  \bibinfo{journal}{Rev. Mod. Phys.} \textbf{\bibinfo{volume}{60}},
  \bibinfo{pages}{537} (\bibinfo{year}{1988}).

\bibitem[{\citenamefont{Paladino
  et~al.}(2002{\natexlab{b}})\citenamefont{Paladino, Faoro, Falci, and
  Fazio}}]{Paladino02}
\bibinfo{author}{\bibfnamefont{E.}~\bibnamefont{Paladino}},
  \bibinfo{author}{\bibfnamefont{L.}~\bibnamefont{Faoro}},
  \bibinfo{author}{\bibfnamefont{G.}~\bibnamefont{Falci}}, \bibnamefont{and}
  \bibinfo{author}{\bibfnamefont{R.}~\bibnamefont{Fazio}},
  \bibinfo{journal}{Phys. Rev. Lett.} \textbf{\bibinfo{volume}{88}},
  \bibinfo{pages}{228304} (\bibinfo{year}{2002}{\natexlab{b}}).

\bibitem[{\citenamefont{Kaulakys et~al.}(2005)\citenamefont{Kaulakys, Gontis,
  and Alaburda}}]{kaulakys}
\bibinfo{author}{\bibfnamefont{B.}~\bibnamefont{Kaulakys}},
  \bibinfo{author}{\bibfnamefont{V.}~\bibnamefont{Gontis}}, \bibnamefont{and}
  \bibinfo{author}{\bibfnamefont{M.}~\bibnamefont{Alaburda}},
  \bibinfo{journal}{Phys. Rev. E} \textbf{\bibinfo{volume}{71}},
  \bibinfo{pages}{051105} (\bibinfo{year}{2005}).

\bibitem[{\citenamefont{Galperin
  et~al.}(2006{\natexlab{b}})\citenamefont{Galperin, Altshuler, Bergli, and
  Shantsev}}]{Galperin06}
\bibinfo{author}{\bibfnamefont{Y.~M.} \bibnamefont{Galperin}},
  \bibinfo{author}{\bibfnamefont{B.~L.} \bibnamefont{Altshuler}},
  \bibinfo{author}{\bibfnamefont{J.}~\bibnamefont{Bergli}}, \bibnamefont{and}
  \bibinfo{author}{\bibfnamefont{D.~V.} \bibnamefont{Shantsev}},
  \bibinfo{journal}{Phys. Rev. Lett.} \textbf{\bibinfo{volume}{96}},
  \bibinfo{pages}{097009} (\bibinfo{year}{2006}{\natexlab{b}}).

\bibitem[{\citenamefont{Saira et~al.}(2007)\citenamefont{Saira, Bergholm,
  Ojanen, and M\"ott\"onen}}]{equdyn}
\bibinfo{author}{\bibfnamefont{O.-P.} \bibnamefont{Saira}},
  \bibinfo{author}{\bibfnamefont{V.}~\bibnamefont{Bergholm}},
  \bibinfo{author}{\bibfnamefont{T.}~\bibnamefont{Ojanen}}, \bibnamefont{and}
  \bibinfo{author}{\bibfnamefont{M.}~\bibnamefont{M\"ott\"onen}},
  \bibinfo{journal}{Phys. Rev. A} \textbf{\bibinfo{volume}{75}},
  \bibinfo{pages}{012308} (\bibinfo{year}{2007}).

\bibitem[{\citenamefont{Kirton and Uren}(1989)}]{kirton}
\bibinfo{author}{\bibfnamefont{M.~J.} \bibnamefont{Kirton}} \bibnamefont{and}
  \bibinfo{author}{\bibfnamefont{M.~J.} \bibnamefont{Uren}},
  \bibinfo{journal}{Advances in Physics} \textbf{\bibinfo{volume}{38}}
  (\bibinfo{year}{1989}).

\bibitem[{\citenamefont{Khaneja et~al.}(2005)\citenamefont{Khaneja, Reiss,
  Kehlet, Schulte-Herbrüggen, and Glaser}}]{grape}
\bibinfo{author}{\bibfnamefont{N.}~\bibnamefont{Khaneja}},
  \bibinfo{author}{\bibfnamefont{T.}~\bibnamefont{Reiss}},
  \bibinfo{author}{\bibfnamefont{C.}~\bibnamefont{Kehlet}},
  \bibinfo{author}{\bibfnamefont{T.}~\bibnamefont{Schulte-Herbrüggen}},
  \bibnamefont{and} \bibinfo{author}{\bibfnamefont{S.~J.}
  \bibnamefont{Glaser}}, \bibinfo{journal}{J. Mag. Res.}
  \textbf{\bibinfo{volume}{172}}, \bibinfo{pages}{296} (\bibinfo{year}{2005}).

\bibitem[{\citenamefont{Cummins and Jones}(2000)}]{cummins2000}
\bibinfo{author}{\bibfnamefont{H.~K.} \bibnamefont{Cummins}} \bibnamefont{and}
  \bibinfo{author}{\bibfnamefont{J.~A.} \bibnamefont{Jones}},
  \bibinfo{journal}{New J. Phys.} \textbf{\bibinfo{volume}{2}},
  \bibinfo{pages}{1} (\bibinfo{year}{2000}).

\bibitem[{\citenamefont{Cummins et~al.}(2003)\citenamefont{Cummins, Llewellyn,
  and Jones}}]{cummins2003}
\bibinfo{author}{\bibfnamefont{H.~K.} \bibnamefont{Cummins}},
  \bibinfo{author}{\bibfnamefont{G.}~\bibnamefont{Llewellyn}},
  \bibnamefont{and} \bibinfo{author}{\bibfnamefont{J.~A.} \bibnamefont{Jones}},
  \bibinfo{journal}{Phys. Rev. A} \textbf{\bibinfo{volume}{67}},
  \bibinfo{pages}{042308} (\bibinfo{year}{2003}).

\bibitem[{\citenamefont{Meiboom and Gill}(1958)}]{meiboom58}
\bibinfo{author}{\bibfnamefont{S.}~\bibnamefont{Meiboom}} \bibnamefont{and}
  \bibinfo{author}{\bibfnamefont{D.}~\bibnamefont{Gill}},
  \bibinfo{journal}{Rev. Sci. Instr.} \textbf{\bibinfo{volume}{29}},
  \bibinfo{pages}{688} (\bibinfo{year}{1958}).

\bibitem[{\citenamefont{Hennessy and Patterson}(2006)}]{Hennessy06}
\bibinfo{author}{\bibfnamefont{J.~L.} \bibnamefont{Hennessy}} \bibnamefont{and}
  \bibinfo{author}{\bibfnamefont{D.~A.} \bibnamefont{Patterson}},
  \emph{\bibinfo{title}{Computer Architecture: A Quantitative Approach}}
  (\bibinfo{publisher}{Morgan Kaufmann}, \bibinfo{year}{2006}).

\end{thebibliography}

\end{document}